\documentclass[10pt,letterpaper]{article}
\usepackage{opex3}
\usepackage{amsmath}
\usepackage{textcomp}
\usepackage{upgreek}

\begin{document}


\title{Two-dimensional disorder for broadband, omnidirectional and polarization-insensitive absorption.}

\author{Matteo Burresi$^{1,2,*}$, Filippo Pratesi$^{1}$, Kevin Vynck$^{1,**}$, Mauro Prasciolu$^{3}$, Massimo Tormen$^{4}$, Diederik S. Wiersma$^{1,2}$}
\address{$^{1,}$European Laboratory for Non-linear Spectroscopy (LENS),
Via N. Carrara, 1. I-50019 Sesto Fiorentino, Firenze, Italy.\\
$^{2}$Istituto Nazionale di Ottica (CNR-INO), Largo Fermi 6, 50125
Firenze, Italy.\\$^{3}$ IOM-CNR, Laboratorio TASC, S.S. 14 Km
163.5, 34149 Trieste, Italy.\\ $^{4}$  Laboratorio Nazionale
TASC-INFM, Basovizza, I-34012 Trieste, Italy.\\$**$
present/current address: Institut Langevin, ESPCI ParisTech, 1 rue
Jussieu, 75005 Paris, France. \email{$^*$burresi@lens.unifi.it}}



\begin{abstract}
The surface of thin-film solar cells can be tailored with photonic
nanostructures to allow light trapping in the absorbing medium.
This in turn increases the optical thickness of the film and thus
enhances their absorption. Such a coherent light trapping is
generally accomplished with deterministic photonic architectures.
Here, we experimentally explore the use of a different
nanostructure, a disordered one, for this purpose. We show that
the disorder-induced modes in the film allow improvements in the
absorption over a broad range of frequencies and impinging angles.
\end{abstract}




\section{Introduction.}
\label{}

The quest for efficient harvesting of solar radiation is one of
the major areas of research in the renewable energy field,
characterized by an interdisciplinary character, ranging from
material science
\cite{schaller_seven_2006,zahler_high_2007,granqvist_transparent_2007,brown_third_2009,chen_polymer_2009,krebs_fabrication_2009}
to optics \cite{andreev_concentrator_2004,Spinelli2012} and
nanophotonics
\cite{Park2009,Han2010,Atwater2010,Ferry2011,meng_absorbing_2011,Mallick2012}.
In particular, the nanophotonic community has been producing a
great deal of alternative strategies to improve the performance of
the various photovoltaic technologies. Among the third-generation
solar cells, the thin-film technologies are the most promising
alternative to the commercially available one, made out of
different, sometimes very expensive and rare, materials (e.g.,
CdTe, CIGS)\cite{brown_third_2009}. Due to the reduced thickness
of these thin films (even below 1 $\upmu$m), nanophotonics is
particularly suited for improving the solar cell absorption
\cite{Yu2010,Callahan2012,Bozzola2012}. Through different photonic
architectures it is possible to augment the optical absorption by
trapping light within thin and ultra-thin films, the latter being
desirable to decrease costs and efficiently extract the
photo-generated charge carriers \cite{kempa_hot_2009}. So far,
nanophotonics has been aiming at increasing the absorption by
patterning deterministic nanostructures (periodic or even
quasi-periodic) on the film, giving rise to partially guided modes
\cite{meng_absorbing_2011,Bozzola2012} in which also the photon
density of states can be manipulated for absorption enhancement
purposes \cite{Callahan2012}. A rising interest in alternative
strategies based on non-deterministic nanostructures has been
developing
\cite{rockstuhl_comparison_2010,martins_engineering_2012,oskooi_partially_2012,kowalczewski_engineering_2012}.
Recently, a properly engineered random patterning which gives rise
to 2D disorder modes on the verge of light (Anderson) localization
\cite{Sigalas1996,vanneste2009,Riboli2011} has been proposed as
light trapping scheme \cite{vynck_disordered_2012}. The impinging
light can easily couple in the 2D random structure increasing the
absorption of the film over a remarkably broad range of
frequencies and angle of incidence \cite{vynck_disordered_2012}.
Given the random nature of these architectures, the optical
properties of it are expected to be less susceptible to
imperfections, suggesting the possibility to envision cheaper
fabrication methodologies.

In this work, we present a proof-of-principle investigation of the
absorption enhancement induced by a 2D disordered photonic
architecture applied to a thin film material. We find that this
strategy improves the absorption of thin films over an extremely
large frequency bandwidth, independently of the angle of incidence
and light polarization. This work provides experimental evidence
that the use of deterministic structures is not the only route for
nanophotonic applications in photovoltaics but rather the
introduction of random photonic architectures can provide
efficient light absorption at any angle of incidence.

\begin{figure}[t]
\centering\includegraphics[width=10cm]{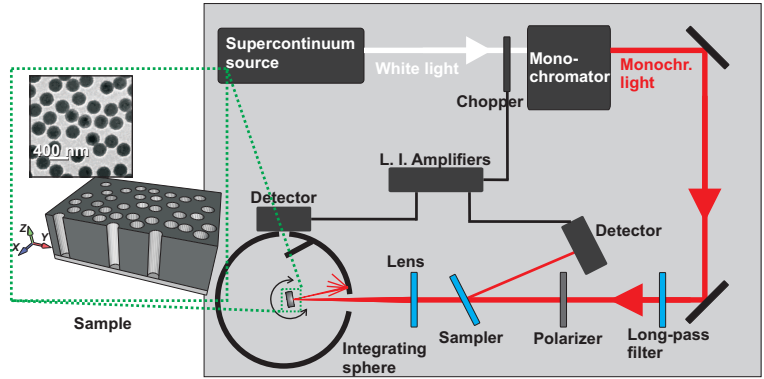}
\caption{Schematic representation of the optical setups employed
to measure the calibrated absorption of the specimens. In the
insets, an electron micrograph of the sample and a 3D sketch of
it.} \label{setup}
\end{figure}

\section{Sample and experimental setup}
\label{}

We investigate a sample made of a layer of amorphous silicon
(a-Si) of thickness $t=930$ nm, deposited on top of 100 nm of
silver. An adhesion layer of chromium (5 nm) is placed between
a-Si and Ag (Supplementary Data). A random distribution (obtained
with a Random Sequential Addition algorithm, as in
\cite{vynck_disordered_2012}) of holes is patterned over an area
of 300 $\upmu$m by 300 $\upmu$m with electron beam lithography and
plasma etching processes. The hole diameter is 270 nm and they
fully perforate the a-Si film (full-etched sample). The hole size
has been roughly chosen to be comparable to the wavelength of
light in the medium in order to have a significant scattering
strength. However, the optimization of this parameter does not
require particular attention since the Mie resonance of a two
dimensional air cylinder in a silicon environment has a very broad
frequency response. This aspect contribute to the optical
robustness of the disorder nanophotonic architecture, which does
not demand for a high quality monodispersity of the holes. The
inset in Figure 1 shows an electron micrograph image of the sample
and a 3D schematics of it. In order to verify the absorption
enhancement not only over a broad spectral range but also for all
angles of incidence and polarizations, an accurate optical setup
for calibrated absorption measurements has been constructed
(Figure 1). The probing white light is emitted by a commercial
supercontinuum source (SM-8-OEM, provided by Leukos SAS). The
laser intensity has been monitored wavelength-by-wavelength to
compensate for any possible spectral drift or power fluctuation of
the laser. As depicted in Figure 1, with the use of a
monochromator we select the probing wavelength before light
impinges on the sample. We insert an optical long pass filter
($@610$nm) to get rid of the higher diffractive orders of the
grating of the monochromator. A polarizer selects the impinging
light polarization and a beam sampler is used to pick up part of
the light at a specific wavelength to monitor the intensity. The
probing light is mildly focused on the random pad with an
achromatic doublet (working distance 200 mm) to reduce the
chromatic aberration and ensure that the focal spot is fully
enclosed in the random pad. To perform measurements as a function
of angle we mount the sample on a rotation stage. Given the
presence of the Ag substrate, the transmission $T$ of the sample
vanishes and thus only reflection ($R$) measurements are required
to obtain the absorption $A=1-R-T=1-R$. More accurately, the
reflection $R$ is constituted by two components, namely the
directly (ballistic) reflected light ($R_o$) and the scattered
light ($R_d$). To measure exactly $R$, the spectrum of the total
reflection must be detected and thus the sample is inserted in an
integrating sphere, as shown in Figure 1. Also, the sample is
tilted by 4 degrees to prevent the directly reflected light to
escape the integrating sphere. For the reference measurements a
silver mirror mounted on an identical mount has been used to
reproduce exactly the experimental conditions. A silicon large
area detector is placed at the surface of the integrating sphere.
Both signals are acquired with a lock-in detection scheme.

\section{Spectroscopic absorption measurements.}
\label{}

We compare the absorption spectrum thus retrieved for the random
pad with the one for a bare slab of the same thickness, as shown
in Figure 2a. Also, the absorption that the bare slab would have
if a perfect antireflection coating (PAR) was present is shown in
green (single-pass absorption). Such a graph has been retrieved by
applying the Lambert-Beer law (neglecting reflectance at the
film/air interface) and by measuring the dispersion of the
material with ellipsometry. We investigate the spectral region
(between 650 nm and 1000 nm) where we expect to have the largest
benefit from the coherent light trapping scheme we propose, since
the absorption drops quickly as a function of wavelengths (see
Figure 2). In this range of wavelengths the absorption length $l$
of a-Si varies approximately from 100 nm to 4 $\upmu$m, which
corresponds to $t/l\approx 9.3$ and $t/l\approx 0.23$,
respectively.
\begin{figure*}[t]
\centering\includegraphics[width=13cm]{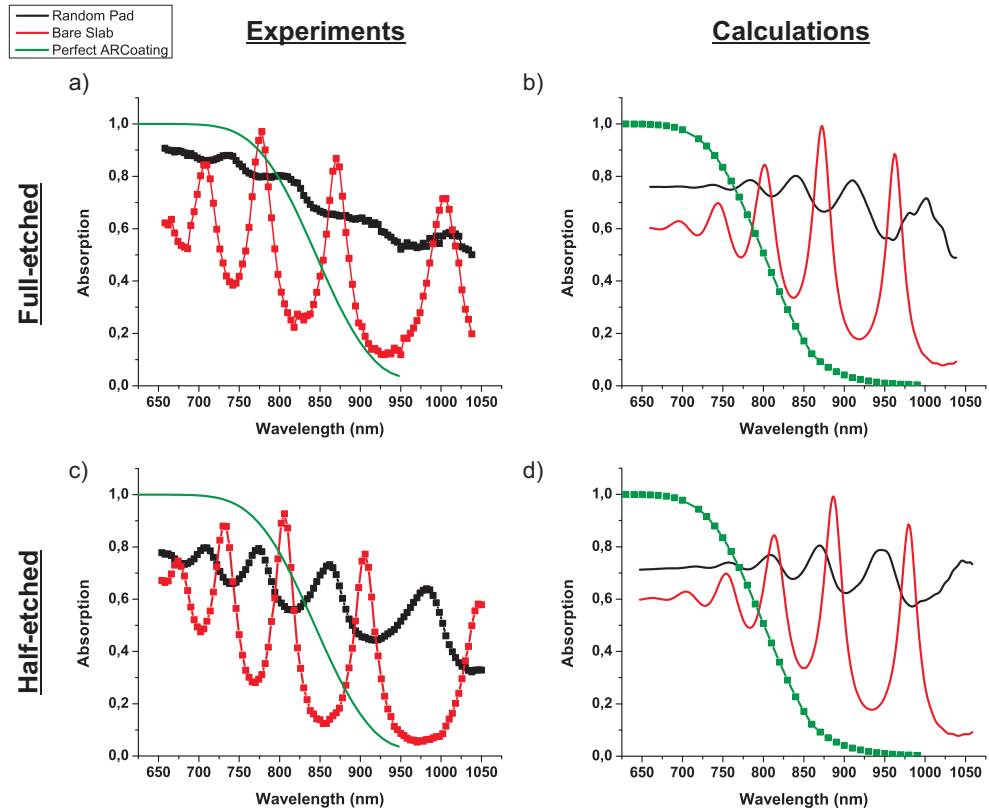} \caption{a) and
b) Measured and calculated absorption for the bare slab and the
full-etched random pad. c) and d) Measured and calculated
absorption for the bare slab and the half-etched random pad. In
green the ideal absorption in absence of surface reflection
(single-pass absorption)} \label{fig2}
\end{figure*}
In the case of the bare slab the Fabry-Perot fringes are visible,
yielding to small frequency windows in the near-infrared where the
absorption is significantly higher than in the case of the PAR.
Due to the optical impedance mismatch at the air/a-Si interface
the absorption performance of the bare slab at shorter wavelength
is far from the PAR case, despite the absorption length of a-Si is
much smaller than the thickness of the slab. In contrast, the
absorption of the random pad at these wavelengths is close to the
PAR case and outperforms both the ideal and the bare slab case for
long wavelengths.

Depending on the absorption length of the bulk material, we can
find different explanations for the absorption enhancement. In
particular, as it has been shown in Ref.
\cite{vynck_disordered_2012}, for a sufficiently weak absorbtion
(long wavelengths regime) the in-plane multiple scattering and
interference effects dictate transport in the structured film.
These give rise to quasi-guided modes in the random pad through
which a relevant absorption enhancement can be achieved.

As sanity check, 3D Finite Difference Time Domain (FDTD)
calculations have been performed using a freely available software
package \cite{Oskooi2010}, by considering the dispersion of the
film obtained by ellipsometry up to 950 nm of wavelength (see
Figure 2b) and a perfect metal instead of the metal substrate. A
good qualitative agreement with the experimental results is found.
Quantitatively, however, the calculated and measured absorption
slightly differs\footnote{The small difference in the frequency of
the resonances between experiment and calculation is mostly given
by the difficulties to implement an the actual dispersion of the
material in the software and the experimental uncertainty on the
thickness of the film.}. It must be pointed out that two important
physical mechanisms are not taken into account by the
calculations: i) the excitation of Surface Plasmon Polaritons
(SPPs) induced by the holes close to the Ag surface and ii) the
presence of dangling bonds \cite{street_hydrogenated_2005} for
a-Si in the holes surfaces. Both phenomena can lead to an increase
of the absorption of the real system, due to the metal absorption
and to the absorption from defect states, respectively. In order
to understand their influence, a different kind of random pad is
fabricated in which the holes are 240 nm shallower than the slab
thickness (half-etched system). Given the distance of the bottom
of the holes to the metal substrate, the excitation of SPPs is
drastically reduced, if not suppressed. In Figure 2c and d
measured and calculated absorption spectra, respectively, are
shown for the half-etched system. The quantitative deviations
between the theoretical expectations and experiment is clearly
reduced. This seems to suggest that the SPPs where mostly
responsible for the quantitative discrepancy between theory and
experiment (Figure 2a and b) and that the presence of dangling
bonds has a minor contribution on the enhancement of optical
absorption. For the sake of completeness, it must be pointed out
that only photocurrent generation experiments could definitely
prove the above statement. However, such a measurement goes beyond
the scope of this work, which aims only to show that with this
novel random nanophotonic architecture a remarkable absorption
enhancement can be achieved. Despite the fact that the absorption
of the random pad does not show evident spectral features, the
enhanced absorption undergoes different regimes. The absorption
mechanism can be better investigated by studying the backscattered
light ($R_d$). $R_d$ is measured rotating the sample such that the
probing beam impinges normally to the surface and the directly
reflected light ($R_o$) escapes from the integrating sphere. In
Figure 3a is shown the measured $R$ and $R_d$, whereas the direct
reflection $R_o$ is retrieved as the difference between the two.
At short wavelengths the reflection $R$ of the random pad is
mostly given by $R_o$, due to the refractive index contrast at the
air/a-Si interface. Since at these wavelengths $R_d$ is only few
percents, we infer that, when light enters the film, it is
absorbed. Indeed the absorption length of the bulk a-Si for
wavelengths between 650 nm to 750 nm is of the order of $10^2$ nm
and thus much shorter than the thickness of the slab and
comparable to the distance between holes. As a result, light is
absorbed long before the in-plane multiple scattering is performed
and thus no photonic mode participates to the increase of the
absorption. Thus, the observed enhancement at short wavelengths is
mostly due to the smaller refractive index contrast at the random
pad/air interface with respect to the bare slab/air one. As a
matter of fact, the introduction of holes in the thin film act as
a 'natural' anti-reflection coating which reduces the
backreflection of light.
\begin{figure*}[t!]
\centering\includegraphics[width=13 cm]{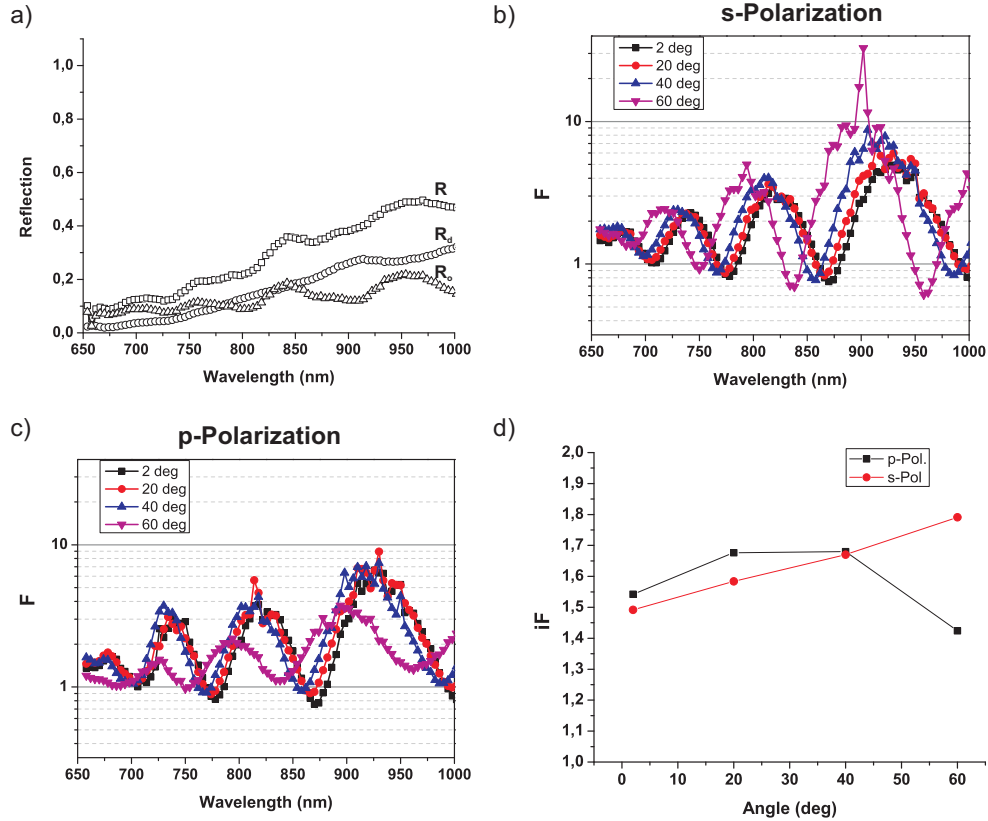} \caption{a)
Total (R, open squares), diffused ($R_d$, open circles) and
directly backscattered reflection ($R_o$, open triangles) for the
full-etched sample. The first is measured a 4 degrees incidence
and the second at 0 degree incidence. The latter has been
retrieved as difference between the previous two. b) and c)
Measured absorption enhancement $F$ for s- and p-polarization,
respectively, for the full-etched sample in log-scale. d)
Integrated (over the probed spectral range) absorption enhancement
$iF$ for s- and p-polarization as a function of angle of
incidence.} \label{fig3}
\end{figure*}
In contrast, at long wavelengths the pronounced increase of
absorption can be explained by coupling to the in-plane disorder
modes. At these wavelengths, the absorption of the randomly
structured film is as large as the absorption of an analogous film
in the single-pass regime but 10 times thicker. The diffused
reflection $R_d$ at longer wavelength, which is around 10\%, is
mostly due to light which undergoes single scattering events
\emph{without} coupling to the slab and then, after reflection by
the metal substrate, propagates through the silicon film. Since
the absorption decreases at long wavelengths, $R_d$ is expected to
increase at long wavelengths.

\section{Angular response of the photonic architecture}
\label{}

For photovoltaic applications, angular and polarization
characterizations are of crucial importance, since the solar
radiation can impinge on a photovoltaic cell under different
angles with random polarizations.

In order to show the performance of a photonic architecture, the
absorption enhancement is the most used figure of merit by the
nanophotonic community \cite{Yu2010,Callahan2012}. It has been
shown that the absorption enhancement achieved by nanophotonic
architectures can have extremely high values (up to $10^2$) in
weakly absorbing materials, in particular when the reference is
the single-pass absorption (green curves in Figure 2)
\cite{Yu2010}. Here we define the spectral absorption enhancement
$F=A_r/A_b$ which compares the spectral absorption of the
nanostructured film $A_r$ with respect to the spectral absorption
of the bare slab $A_b$. We performed measurements rotating the
sample such that the angle of incidence varies from 4 to 60
degrees, for both slab and random pads. Figure 3b and c show $F$
as a function of angle for two orthogonal polarizations (s and p,
respectively). Given the pronounced  Fabry-Perot oscillations,
frequency windows where the absorption enhancement is as high as
10 times are visible, with a clear improving trend towards long
wavelengths, where the absorption length of the material is
longer.

The overall trend of $F$ as a function of angle and wavelengths
can be better seen by calculating the integrated $iF$ weighted
with the air-mass coefficient 1.5 spectrum $I$ for the incident
light. $iF$ is defined as
$iF=\frac{\int_{\lambda_1}^{\lambda_2}A_r I
d\lambda}{\int_{\lambda_1}^{\lambda_2}A_s I d\lambda}$, where, in
our case, $\lambda_1= 650$ nm and $\lambda_2=1000$ nm. Figure 3d
shows $iF$ as a function of incident angle for the two
polarizations. As expected both polarizations exhibit a similar
$iF\approx 1.6$ at small angles. Please note that given the
significant absorption of the material with respect to its
thickness, the architecture cannot achieve orders of magnitude of
absorption enhancement \cite{yu_thermodynamic_2012}.
Quantitatively, the response as a function of angle is similar for
both polarizations, as expected given the random nature of the
nanostructures, but a clear difference trend can be seen. In
particular for the p-polarization $iF$ slightly decreases at large
angles. As the angle of incidence approaches the Brewster's angle,
the p-polarization experiences a smaller reflectance which in turn
increases the absorption of the film. Experimentally we verified
that the absorption of the bare slab increases faster than
nanostructured case as a function of angles, yielding a decrease
of $iF$.

\section{Conclusions}

In conclusion, we experimentally verified that the presence of
in-plane disorder modes in a thin film can be used to
significantly enhance the light absorption of commercially
available thin-film solar cells. Such an increase of absorption
occurs on a broad frequency bandwidth and angular range (up to 60
degrees of incident angle). On one hand, at long wavelengths we
obtained an increase of the absorption due to the light coupling
to the modes arising from the in-plane multiple scattering and
interference effects. On the other hand, at shorter wavelengths
the absorption enhancement is given by the lower effective
refractive index induced by the nanopatterning, which acts as a
broadband antireflection coating embedded in the film itself. In
this wavelength regime the very short absorption length does not
allow the guided modes formation by multiple scattering. We
believe that the omnidirectionality of the absorption enhancement
for all polarizations makes this disorder photonic strategy
particularly promising for thin-film photovoltaic applications.
This work experimentally proves that deterministic structures are
not the only possible way towards an improvement of thin-film
absorption by nanophotonic means.

\section*{Acknowledgement} We wish to thank F. Riboli and G. Conley
for fruitful discussions. This work is supported by the European
Network of Excellence Nanophotonics for Energy Efficiency,
CNR-EFOR, and ENI S.p.A. Novara.
%
%

\end{document}